# Mode-Locked Topological Insulator Laser Utilizing Synthetic Dimensions

Zhaoju Yang[*], Eran Lustig[*], Gal Harari, Yonatan Plotnik, Miguel A. Bandres, Yaakov Lumer, Mordechai Segev[†]

*Physics Department and Solid State Institute,*

*Technion– Israel Institute of Technology, Haifa 32000, Israel*

*equal contribution

†Corresponding author. Email: msegev@tx.technion.ac.il

**Abstract**

We propose a system that exploits the fundamental features of topological photonics and synthetic dimensions to force many semiconductor laser resonators to synchronize, mutually lock, and under suitable modulation emit a train of transform-limited mode-locked pulses. These lasers exploit the Floquet topological edge states in a 1D array of ring resonators, which corresponds to a 2D topological system with one spatial dimension and one synthetic frequency dimension. We show that the lasing state of the multi-element laser system possesses the distinct characteristics of spatial topological edge states while exhibiting topologically protected transport. The topological synthetic-space edge mode imposes a constant-phase difference between the multi-frequency modes on the edges, and together with modulation of the individual elements forces the ensemble of resonators to mode-lock and emit short pulses, robust to disorder in the multi-resonator system. Our results offer a proof-of-concept mechanism to actively mode-lock a laser diode array of many lasing elements, which is otherwise extremely difficult due to the presence of many spatial modes of the array. The topological synthetic-space concepts proposed here offer an avenue to overcome this major technological challenge, and open new opportunities in laser physics.

Topological insulators [1,2] are a new phase of matter with an insulating bulk that supports topologically protected unidirectional edge states. These edge states are endowed with topological invariants and their transport is therefore robust against defects and disorder as long as the topological bandgap is preserved. After their observation in electronic systems [3], topological insulators have been studied in many physical systems such as photonics [4–14], acoustics [15] and mechanics [16]. Outside condensed matter, topological insulators were first proposed to occur also with electromagnetic (EM) waves [4,5], and shortly thereafter observed with microwaves in gyro-optic materials [7]. At that point it was clear that introducing the topological concepts into photonics would lead to many new discoveries. However, this transition turned out to be difficult because at optical frequencies the gyro-optic response to magnetic fields is extremely weak, resulting in only a tiny topological bandgap that would be robust only against extremely weak perturbations. Thus, the realization of photonic topological insulators had to rely on artificial magnetic fields [17], rather than on magneto-optics. In the following years, several ideas were proposed along these lines [8,10,13,18,19], until eventually, photonic topological insulators were demonstrated experimentally in 2013 [11,12]. Both of these systems relied on photonic crystal settings incorporating artificial gauge fields. However, topological insulators do not necessarily have to rely on spatial gauge fields. Rather, topological phenomena can be constructed in synthetic dimensions [20–22], using internal degrees of freedom such as spin in atomic systems [23,24], or by using external degrees of freedom such as shaking harmonic traps containing cold atoms [25]. In photonics, synthetic dimensions were proposed for generating solitons in higher dimensions [26], and were demonstrated in the context of topological pumping [27,28]. In this vein, topological insulators were proposed to occur in photonic platforms utilizing synthetic space, such as the modal dimension of a resonator [29,30], with intriguing suggestions for observing high-dimensional quantum Hall effect and Weyl points [29,31] by controlling the phase of the dynamic modulation [9]. Eventually, in the past year, photonic topological insulators in synthetic dimensions were demonstrated experimentally, by utilizing a lattice of spatially-modulated waveguides [32].

The unique features of topological photonics have led to many new applications where robustness is important, such as delay lines [12], or more recently, topological insulator lasers [33–35], which includes gain within the laser cavity (edges) and utilizes the topological edge states as the lasing mode. Topological insulator lasers combine the concepts of topological insulators with the fundamentals of lasers, which introduce gain, loss, and lasing action in a topological platform. Specifically, in the presence of an artificial gauge field [34], an optical lattice consisting of resonators exhibits topological edge states. Incorporating gain on the edges of the 2D lattice of coupled resonators causes the topological lasing state to dominant and extend over the perimeter of the laser cavity, resulting in unidirectional energy flux, high slope efficiency and single-mode lasing even high above the laser threshold [27,28]. This conceptual advance may also lead to new designs for masers [36], phonon lasers [37], spacers [38–40], and to lasing action in the recently discovered exciton-polariton topological insulators [41].

Here, we take the concept of topological insulator laser to the next level, and propose multi-element mode-locked topological insulator lasers in synthetic dimensions.

Let us first explain the technological and conceptual challenge to mode-lock a laser diode array. An array of evanescently-coupled laser diodes was proposed 35 years ago as an avenue to achieve a coherent high-power laser source based on semiconductor technology [42,43]. However, within a decade it became clear that semiconductor laser arrays tend to lase in many modes simultaneously, and that their field pattern and spectra vary considerably even under small variations in the gain, and that the mutual coherence of the array is greatly reduced with increasing the number of laser elements. Consequently, current technology of laser arrays is limited to acting as a pump for solid state lasers [44], but not as a high-power coherent laser source, as was originally envisioned. Another major consequence of the extreme difficulty in locking multiple elements of semiconductor lasers is the inability to synchronize them, that is, to generate mode-locked pulses from a laser diode array. The difficulty arises from the fact that a laser diode array has multiple spatial modes (array modes), and each spatial mode has a set of associated resonance frequencies that are shifted from those of the other array modes. In addition, the spectra of modulated laser diodes is fundamentally chirped (i.e., the spectrum varies during the modulation) [45,46] and the chirp of different laser elements is sensitive to small fabrication differences. On the other hand, achieving mode-locking requires synchronizing all the lasing frequencies of all the lasing elements, which becomes extremely hard when the laser has multiple spatial modes, as all laser arrays fundamentally have. For this reason, the only attempts to mode-lock a laser diode array relied on having external cavities [47–49], and even those were left unfollowed, never maturing into real technology.

In contrast to the early attempts to actively mode-lock laser diode arrays, we propose a system that exploits the fundamental features of topological photonics to force many semiconductor laser resonators to synchronize, mutually lock, and emit a train of transform-limited mode-locked pulses. Our proposed system relies on a topological platform that utilizes a Floquet topological edge state in a 2D synthetic space that is comprised of one modal dimension (the modes of the resonators) and one spatial dimension (the array of resonators). The edge state therefore lives on four edges: right/left edge (rightmost/leftmost ring, all modes) and the bottom/upper edge (lowermost/uppermost mode, all rings). The "interior region" of our synthetic space topological insulator laser is comprised of all the modes that are enclosed by the edgestates of our system. For example, if we induce gain on modes 1 and N, then the interior of our topological insulator is comprised of all the modes 2 to N-1 of all the rings except the side rings. The Chern number in our synthetic space system is alternating with the modulation frequency, but in our parameters range we use the first bandgap which sets the Chern number to 1. The fundamental impact of the synthetic dimension here lies in the fact that one of its dimensions is a ladder of equally spaced modes (frequencies). Thus, unlike in a real-space topological insulator laser (which can lase at a single frequency), here the spectrum of the a topological edge-states is a frequency comb. This setting, together with gain saturation (present in all lasers), selects a single lasing mode – ***the Floquet topological edge mode - for the entire array, with which has a <u>single</u> set of equally-spaced resonance frequencies***. This unidirectionally-propagating lasing state in synthetic dimensions is robust to imperfections and small variations in the individual laser elements. It induces a fixed phase-difference between the modes of the cavity and therefore generates constructive interference between these modes, causing the laser light to be produced as transform-limited mode-locked pulses.

To construct the topological laser in synthetic dimensions, we adopt the modulated ring resonators [29,50] as our building block. In the absence of group velocity dispersion in the underlying ambient medium, each ring resonator supports a discrete set of resonant modes at equally-spaced frequencies described by $\omega_m = \omega_0 + m\Omega$, where $\Omega$ is the free spectral range of the ring. We assume static coupling only between modes with the same frequency at the nearest neighbor resonators, with a coupling strength $t_0$. In addition, each resonator is modulated with the frequency $\Omega$, which induces dynamic coupling (coupling strength $t_f$) between modes separated by $\Omega$. Combined with the mode dimension, the tight-binding Hamiltonian [30] of the system can be expressed as

$$H = \sum_{j,m} \omega_m a_{j,m}^\dagger a_{j,m} + \sum_{j,m} t_0 (a_{j,m}^\dagger a_{j+1,m} + a_{j+1,m}^\dagger a_{j,m}) + \sum_{j,m} 2t_f \cos(\Omega t + \phi_j)(a_{j,m}^\dagger a_{j,m+1} + a_{j,m+1}^\dagger a_{j,m}) \quad (1)$$

where $a_{j,m}$ ($a_{j,m}^\dagger$) is the annihilation (creation) operator at $j$th site for mode $m$, $t$ is time and $\phi_j$ is the modulation phase at the $j$th resonator. If we set the modulation phase to be $\phi_j = x\phi$ ($x$ is the unitless position of the resonator where each two neighboring resonators are separated by a distance 1), under the condition of $t_f \ll \Omega$ and the rotating wave approximation, the dynamics of this photonic system can be mapped to the Hofstadter model [51] in a space with one spatial dimension and one modal dimension. ). Considering a strip geometry (finite number of rings and infinite modes in frequency dimension), we prove the existence of a topological band structure and topological edge states, as shown in Fig. 6(a) [Appendix A].

Our design for the mode-locked topological laser system in synthetic dimensions is made up of identical modules, connected in a way that maintains their topological synthetic-space features. We first define the single module and then describe the full laser array system. The single module consists of two rows of evanescently-coupled resonators (Fig. 1a). The rows are decoupled from one another, but they are connected through the two side resonators. In total there are N resonators, each of which supports M equally-spaced modes which can interact with light within the bandwidth of their active medium. The two-dimensional shaded regions above and below the rows describe the synthetic space, which includes the position and the mode number. In the absence of gain and loss, this system displays a topological phase when $\phi \neq 0, \pi$, whereas for $\phi = 0, \pi$ the system is topologically trivial. The maximum topological bandgap corresponds to $\phi = \pi/2$. In the topological phase, this system exhibits unidirectional edge states spanning across the bandgaps. To facilitate lasing of the edge state, we introduce optical gain and loss in the synthetic space, in the positions marked by blue and grey circles, respectively. The evolution of the field in this laser system is governed [33] by

$$i\partial_t \Psi = H\Psi + \left(\frac{ig\mathbb{P}}{1+\frac{\Sigma_m|\varphi_m|^2}{I_{sat}}} - i\gamma\right)\Psi \qquad (2)$$

where $H$ indicates the Hamiltonian of the setup shown in Fig. 1(a) and $\Psi$ is the column field vector for the amplitudes of all the relevant modes in the resonators ($\varphi_m$ indicates the laser field in m-th synthetic mode). The second term includes the gain and linear loss γ in each resonator, which manifests the fact that the lasers are fundamentally non-Hermitian and nonlinear entities. The optical gain via stimulated emission is inherently saturable ($I_{sat}$), and the summation is taken over all modes of each resonator. Here, $\mathbb{p}$ stands for the spatial profile of the pump. The gain ($g$) is provided only to the modes on the perimeter of the synthetic space [blue circles in Fig. 1(a)]. Here the gain profile is assumed to be broadband and does not depend on the modulation. Note that, since the gain saturation intensity for each resonator is the same and the edge state in synthetic dimensions occupies multiple modes of the side resonator, we need to accordingly pump the side resonators harder for smoothening the nonlinearity-induced potential barrier between side resonator and its neighbors. We optimize and set the gain of side resonators M/2 times larger than that of the bulk resonators through parameter sweeping.

The dynamics in this array of resonators is simulated by solving Eq. (2) numerically starting from random amplitudes and phases. As shown in Fig. 1(b), the topological lasing state ($\phi = \pi/2$) localizes at the edges of the synthetic space, which amounts to populating all the modes of the two side resonators, as well as the lowermost mode of the first row and the uppermost mode of the second row. Accordingly, we pump these modes in the pertinent resonators to obtain maximum overlap of the gain with the edge modes. In real space, the light in the lasing state propagates through all the resonators and saturates the gain (Fig. 7). In the synthetic space picture, this lasing state corresponds to the topological edge state on the perimeter of a 2D lattice with a perpendicular magnetic field. By contrast, in the corresponding topologically-trivial system (same system with $\phi = 0$), the lasing state stays almost stationary and penetrates deeper into the lossy bulk, as shown in Fig. 7(c), which leads to lower lasing efficiency. Figure 2(a) displays the simulated laser output versus pumping intensity, with the blue (grey) describing the operation of the topological nontrivial (trivial) laser. The slope efficiency of the topological laser is about twice that of the trivial laser.

Endowed with the topological properties, we expect the lasing state to exhibit robust transport associated with topological edge states. To examine this, we introduce fabrication disorder into the laser system, in the form of random variations in the resonant frequencies expressed as on-site disorder in the tight-binding model (with magnitude in units of the coupling coefficient). As shown in Fig. 2(b), in the presence of disorder, the topological laser still operates in the fashion of one-way edge state and the slope efficiency of the topological laser stays at the high level. The slope efficiency tends to approach that of the trivial laser even when the disorder level is strong enough to close the topological band gap. Here, we assume that the disorder is randomly distributed to all the modes in the synthetic space, as described by a tight-binding theory.

The topological edge state propagating in the edges of synthetic space indicates that the light of the laser occupies all the modes in the side resonators, as shown in the yellow shaded regions in Fig. 1(b). The calculated spectra for the topological and trivial lasers under the same level of pumping are marked by blue and grey curves in Fig. 2(c) and 2(d), respectively. Since the topological laser lases in a single Floquet mode unlike the trivial laser that lases in many Flouqet modes, the spectrum of the topological laser shows higher and smoother spectrum intensity distributions than that of the trivial laser, which indicates that the light in the topological laser system evolves through all the (pumped) modes in the synthetic dimension, whereas the light becomes more localized in the frequency dimension for the trivial laser, populating only a subgroup of modes. Furthermore, as shown in the right part of the panel 2c, the topological laser guarantees single Floquet mode lasing within the topological band gap in the synthetic space, leading to the form of a frequency comb with spacing Ω. On the other hand, in the trivial system lasing is weaker and occurs at multiple Floquet modes, as shown in the right part of the panel d.

The power flux of the light in the topological cavity is unidirectional. First, as shown in Fig. 1(b), the unidirectionality can be observed by noticing that the power increases while traveling counter-clockwise around the topological cavity in synthetic space, and then abruptly drops after coupling to the neighboring resonator. This can also be observed from the spectrum [Fig. 2(c)]. Second, the dispersion relation of the single one-way edge state with the quasi-energy within the band gap makes the topological edge state in synthetic space correspond to a single-valued wavevector in frequency dimension. This single-valued wavevector determines the fixed phase-difference between the adjacent modes of side resonator [e.g., $0.3\pi$, green dot in Fig. 6 corresponding to Fig. 3(a)]. Figure 3(a) shows the result of time-domain simulations of such evolution (with a fixed phase between adjacent modes) for a duration of $800\tau$ ($\tau = 2\pi/\Omega$), when the topological laser system arrives at steady state. On the other hand, as shown in Fig. 3(b), in the topologically-trivial laser system the phase difference between the modes exhibits multiple values and the power in each is weak, which, unlike the topological case, lacks a single dominant peak. Evidently, in the trivial system the spatial and temporal modes cannot be synchronized to be all in-phase periodically in time.

The fixed phase difference between adjacent modes [Fig. 3(a)] hints that the laser operates in a mode-locked fashion. Essentially, the side resonators of the topological module generate a frequency comb with a fixed-phase difference between the modes, which give rise to the topologically protected phase-locking of the light in all the ring resonators in the system, including the side resonators which lase in multiple modes simultaneously. The lasing modes constructively interfere and produce a periodic train of narrow pulses [Fig 4(a)]. In fact, this mechanism gives rise to active mode-locking of the entire laser array system. Figure 4(a) shows the pulses, each of width (at FWHM) of ~$0.045\tau$, separated by $\tau = 2\pi/\Omega$ emitted from the topological lasing system in synthetic dimensions. We find that the topological mode-locked laser generating pulses is immune to the disorder on the ring resonators, as presented in Fig. 4(b). In contrast to that, the trivial system exhibits multi-mode lasing and backscattering, which does not produce clean mode-locked pulses in the absence [Fig. 4(c)] and presence [Fig. 4(d)] of disorder.

Having described a single module of our system, consider now an assembly of modules, as described by Fig. 5(a). This is an array of $n$ modules, with two sets of $n$ side rings whose modes are locked at constant phases to yield mode-locked pulses from the entire laser diode array. Figure 5(b) presents the intensity of the light emitted from one set of the side resonators as they constructively interfere due to the topological lasing process. Thus, the topological laser array generates a train of clean mode-locked pulses even in the presence of considerable disorder [Fig, 5(d)]. The trivial laser array emits fluctuating pulses with much less output power [Figs. 5(c, e)] since it cannot constructively interfere at synchronized times, as the topological laser. The pulse emitting efficiency in the topological device is about 10 times higher than that of the trivial laser. These results are further confirmed by verifying that the peak power scales quadratically with the number of emitters, while – had it been incoherent – the peak power would be merely the sum of the time-averaged intensities, which of course would never display clean pulses. Figure 5g shows exactly this, and serves as an unequivocal proof that the topological platform can indeed force mode-locking of very many semiconductor laser emitters – exactly because it forces a specific (FIXED) phase difference between adjacent emitters, whereas for the trivial platform the relative phase between emitters varies fast with time. As such, the light emitted from the topological system interferes coherently at round-trip times with peak power scaling quadratically with the number of emitters - leading to a train of clean mode-locked pulses, while the light from the trivial system cannot do that because the different emitters are not synchronized in their phases.

This naturally raises questions about the reasons: what is it in the topological features that makes the ensemble of semiconductor laser elements mode-locked? The answer is two-fold. First, the topological edge mode always moves at a constant velocity, which not only forces all the lasers to lock in their set of resonance frequencies, but also gives rise to a constant phase between modes in adjacent resonators. Effectively, the topological landscape makes sure that the system lases in a single Floquet edge mode and the dynamics in the synthetic dimension forces all the resonant frequencies to lock, giving rise to mode-locked pulses. Second, the topological protection of transport endows the system with robustness against disorder. Without the topological features, the assembly lases in multiple spatial modes which cannot synchronize their resonant frequencies, and hence cannot mode-lock.

The system we propose here can be experimentally realized using ring resonators made of InGaAsP quantum wells similar to the rings in Ref. [34]. The synthetic dimension can be constructed in the rings by using electro-optic modulators [52,53] modulating the refractive index. To match the modulation rate to the spacing between adjacent modes of the ring, the ring radius can be about 200 μm for a modulation rate of 80GHz, which is accessible with current technology [52]. The relation between the radius and modulation rate is determined by $n_{eff}\frac{\Omega}{c}2\pi r = 2\pi$. This would correspond to ~0.6ps pulses separated by $\tau = \frac{2\pi}{\Omega} = 12.5$ps. Increasing the number of modes would make the pulses narrower, with the same repetition rate. To facilitate the lasing of the artificial edge and selectively populate the target mode, we can adopt a general design of periodically arranging Cr-Ge structures on top of the resonator along the azimuthal direction [54,55]. The specific mode experiencing maximum gain will reach the lasing threshold first and lase in a single frequency. Therefore, we can provide azimuthal gain/loss modulation of a smaller azimuthal order (e.g. 30) to the rings in the lower chain [as in Fig. 1(a)], and modulation with higher azimuthal order (say, 50) to the rings in the upper chain. In this design relying on synthetic-space topology, we combine the upper edge (at mode 50) and the lower edge (at mode 30) in a desirable fashion for active mode-locking of a large number of semiconductor laser elements.

To summarize, we utilized the principles of mode-locked lasers combined with the concept of photonic topological insulators in synthetic dimensions to propose a system that can force a large array of semiconductor laser to emit transform-limited mode-locked pulses. While the design here is suggested as a proof of concept, it can be optimized to a specific laser system. At the heart of the ideas is the unique feature of topological edge state that forces a fixed phase difference between adjacent sides as it evolves, and the topological protection which provides robustness to disorder and imperfections. Our system here is topological in synthetic space, hence the single Floquet topological edge state in synthetic space forces a fixed-phase difference between the multiple modes in side resonators and locks all the elements in the laser array. As a result, the comparison between topological and non-topological structures is dramatic: the topological system emits a train of clean transform-limited pulses whereas the trivial system emits irregular bursts characteristic of modulated multi-element lasers that lack locking. In the proposed design we adopted the highest or lowest mode as the artificial edge in the two rows of the rings, but this is not a necessary condition. As we show in Appendix F, such a system emits mode-locked pulses even when the 'edge' is surrounded by a lossy bulk. The reason is that the nonlinear gain/loss automatically introduces a "potential barrier" resulting in the formation of an effective interface separating two regions [Fig. 11 in Appendix F]. Also, as we show in Fig. 13 there, the small side lobes between the pulse peaks can be further suppressed by including inhomogeneous broadening. Other than emitting mode-locked pulses, we also show that a similar topological laser array system, where the laser utilizes the gain from all the ring resonators, can be designed to have single-frequency lasing in synthetic dimensions [See details in Appendix. I]. We envision many more possibilities relying on utilizing topological phenomena in synthetic dimensions in lasers and amplifiers, suggesting a new route for developing active photonic devices.

This work was sponsored by the Israel Science Foundation, and by an Advanced Grant from the European Research Council.

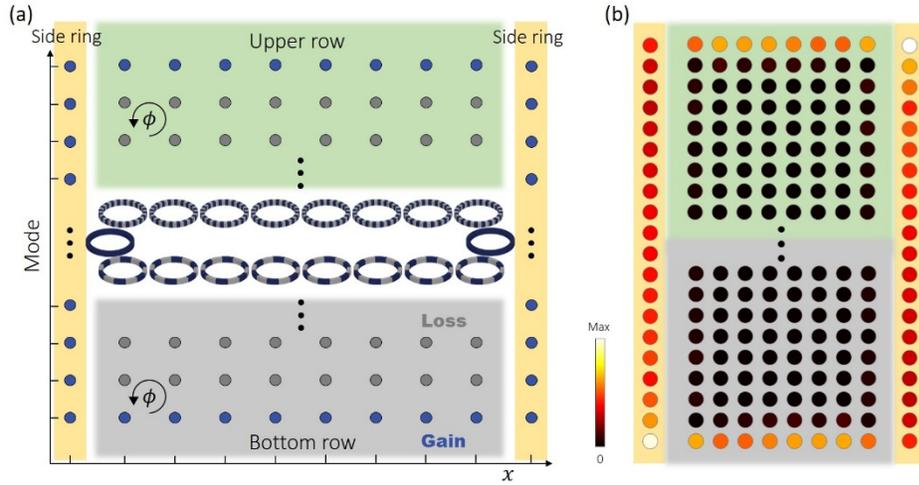

Figure 1.

Schematic of one module of the topological insulator laser in synthetic dimensions. (a) The photonic array consists of modulated rings, in the form of two independent rows connected by two side resonators. The blue-white staggered pattern indicates the different pumping schemes. The shaded regions show the synthetic space including one spatial and one synthetic mode dimensions. The blue (grey) circles indicate the gain (loss), where the gain is provided only to the upper (lower) mode in the upper (lower) row, and to all modes in the side rings. (b) The typical topological lasing state extending over the artificial perimeter of the 2D synthetic space. The color bar indicates the intensity of light. The unitless parameters for simulation are N=18, M=20, $t_0 = t_f = 1$, $\Omega = 100$, $I_{sat} = 1$ and $\gamma = 0.3$. The topological system has bandgap size of $1.6t_0$.

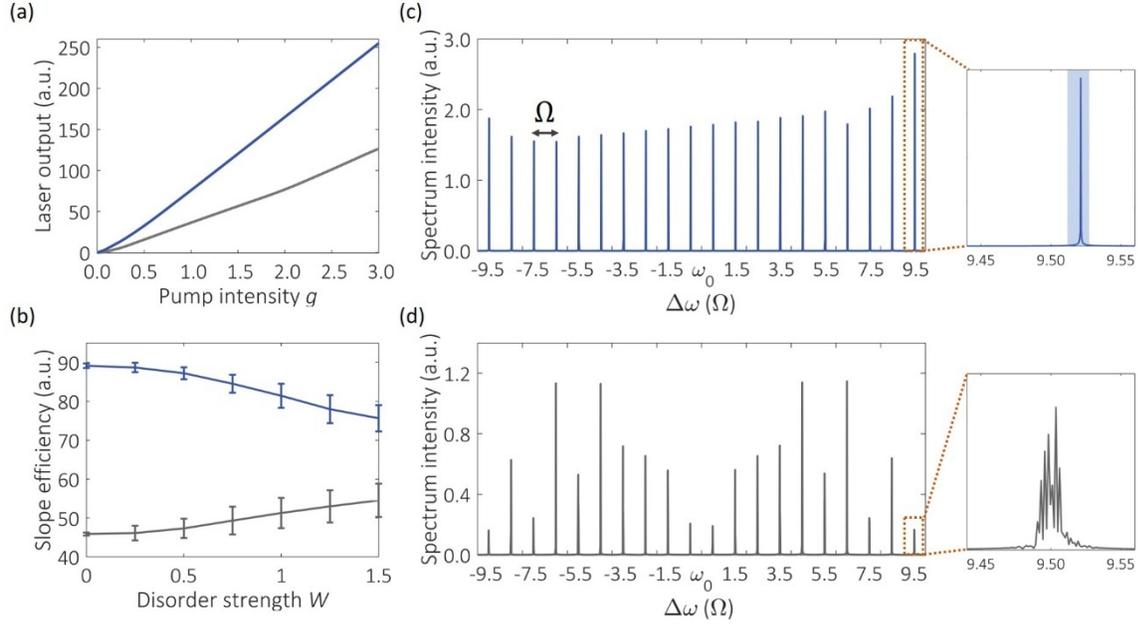

Figure 2

Laser output versus pumping intensity and the spectra of the topologically nontrivial and trivial lasers. (a) The total output power versus pumping intensity. The slope efficiency of the topological laser is about twice that of the trivial laser. (b) Slope efficiency versus disorder strength. The on-site disorder comprises of independent random variations in the resonance frequencies, uniformly distributed in [-W, W], where W is in the units of $t_0$. This panel presents the mean of 100 realizations of disorder for every point of disorder strength. (c-d) Frequency spectra of the topologically nontrivial (c) and trivial (d) lasers, as emitted from the connecting side resonator on the right-hand side; the panels on the right show the enlarged spectra around 9.5Ω. The topological system emits a comb of narrow single frequencies, whereas the trivial system has multiple modes around each central frequency. The original resonances in each ring are set to be $-\frac{M-1}{2}\Omega, \ldots \frac{M-1}{2}\Omega$ centered at $\omega_0$ ($\Omega = 100$ is the free spectral range, M = 20 is the number of modes in each ring; $\omega_0$ is set to 0 for simplicity). The topological landscape causes a constant shift to all the resonance frequencies by a value of 1.9, as shown in the enlarged spectrum in (c) (see explanation in Appendix A). The blue and grey curves indicate the results of represent the topological and trivial lasers, respectively. The blue shaded region in the right of panel (c) indicates the topological bandgap.

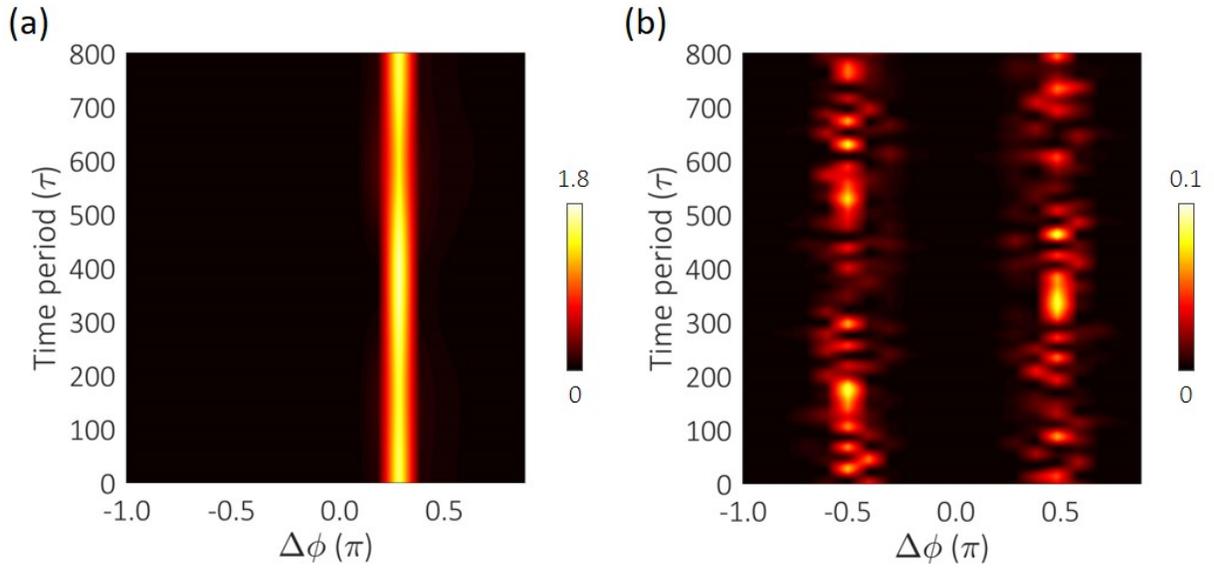

Figure 3

Temporal dynamics of the phase difference between adjacent resonant modes in the side rings. (a) The topological laser exhibits a fixed-phase difference between the modes in the right-side ring, which gives rise to the mode-locking. (b) The trivial laser displays multi-valued phase differences between the modes of the same ring, which therefore cannot be synchronized to yield mode-locked operation. The color bar indicates the intensity of the field with a specific phase difference. The time axis is in units of $\tau = 2\pi/\Omega$.

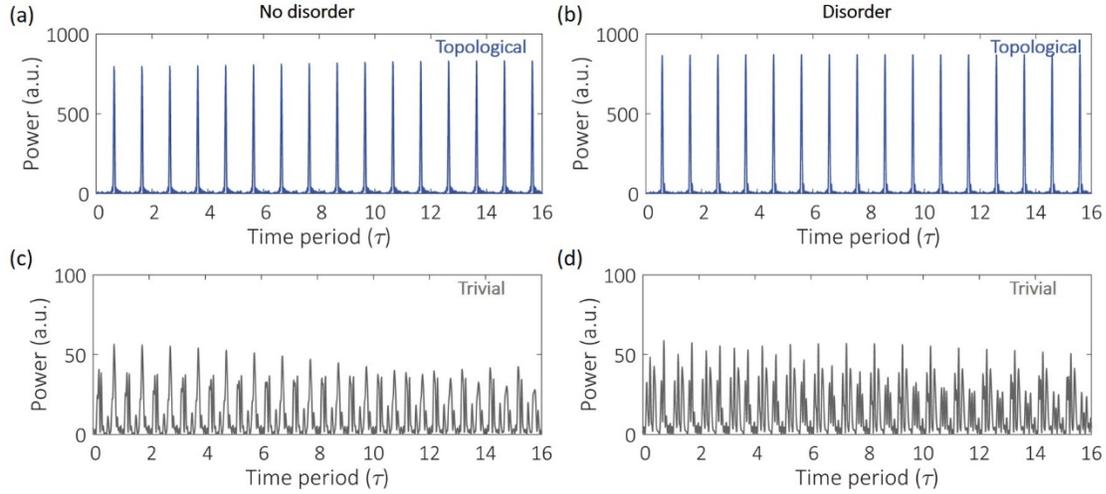

Figure 4

Temporal dynamics of the laser output emitted from the side resonators. The topological module emits mode-locked pulses (a) and maintains clean mode-locked operation even under considerable disorder with W=0.5 (b). On the other hand, the trivial laser system ($\phi = 0$) displays multiple bursts of power, which broaden and fluctuate in the absence (c) and presence (d) of on-site disorder which introduces frequency mismatch among the resonators. The blue and grey curves indicate the results of topological and trivial lasers, respectively. The time axis is in units of $\tau = 2\pi/\Omega$. Note that in the simulations, background noise is present at all times. The background noise is distributed in $[-\delta, \delta]$ and $\delta = 0.2$.

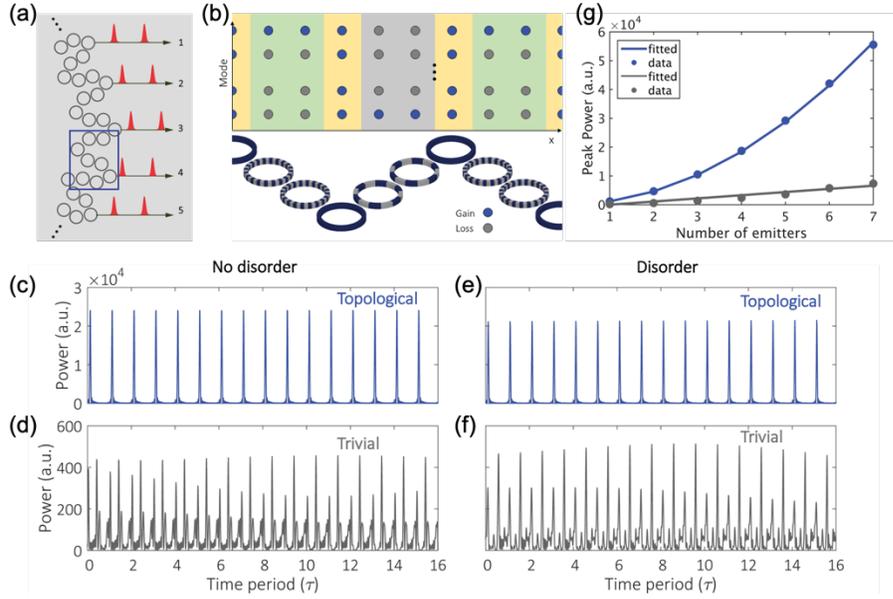

Figure 5

Topological mode-locked laser array. (a) Scheme showing a laser array of n=5 modules [each of which described by Fig. 1(a)]. The side resonators in the laser array configuration connect different chains of ring resonators. The lower panel shows an enlarged view of blue box in panel (a). The laser light is extracted through output couplers. (b) The gain profile in synthetic dimensions. The ring resonators correspond to the ones within the blue box in panel (a). (c, e) The topological laser array generates a train of mode-locked pulses (c) which are emitted from the top (or bottom) connecting rings, and remain clean mode-locked pulses even in the presence of considerable on-site disorder with W=0.3 (which introduces frequency mismatch among the resonators), (e). (d, f) The trivial laser array emits fluctuating pulses with much less output power even in the absence (d) and presence (f) of disorder. (g) Peak power scales with the number of emitters quadratically for topological laser (blue) and linearly for trivial laser (grey). The simulation parameters we used are the same as those in Fig. 1(a). The time axis is in units of $\tau = 2\pi/\Omega$. In panels (c-f) the laser output power is calculated from the sum of the electric fields of side rings in the laser array.